\PassOptionsToPackage{table}{xcolor}
\documentclass{article}
\usepackage[T1]{fontenc}
\usepackage{iclr2027_conference,times}

\usepackage{amsmath,amsfonts,bm}

\def\eqref#1{equation~\ref{#1}}

\def\1{\bm{1}}

\DeclareMathAlphabet{\mathsfit}{\encodingdefault}{\sfdefault}{m}{sl}
\SetMathAlphabet{\mathsfit}{bold}{\encodingdefault}{\sfdefault}{bx}{n}

\usepackage[table]{xcolor}
\usepackage{pifont}
\usepackage{booktabs}
\usepackage{graphicx}
\usepackage{multirow}
\usepackage{tabularx}
\usepackage{hyperref}
\usepackage{microtype}
\usepackage{url}
\usepackage{fancyvrb}
\usepackage{tcolorbox}
\usepackage{amssymb} 
\newenvironment{revision}{\begingroup}{\endgroup}
\tcbuselibrary{skins,breakable}
\usepackage[scaled=0.88]{inconsolata}
\definecolor{GRODERow}{RGB}{235,243,250}
\newif\ifrevisionred
\revisionredtrue
\definecolor{RevisionRed}{RGB}{190,0,0}
\newcommand{\rev}[1]

\title{GRODE-RTL: Source-Mapped Dependency Views for LLM-Based RTL Repair}

\author{
  Jingyi Zhou$^{1}$ Yi Liu$^{1}$, Bin Sun$^{2}$, Ziyang Zheng$^{1}$, Hongji Zhang$^{1}$, Dongsheng Zuo$^{1}$, Qiang Xu$^{1}$\thanks{Correspondence to: Qiang Xu \texttt{qxu@cse.cuhk.edu.hk}.} \\
  $^{1}$Department of Computer Science and Engineering, The Chinese University of Hong Kong \\
  $^{2}$Institute of Computing Technology, CAS \\
}
\iclrfinalcopy

\begin{document}

\maketitle

\begin{abstract}
Large Language Models show great potential for automated Register-Transfer Level (RTL) repair, yet they struggle with non-local faults where root causes and failing outputs reside far apart in source code. Treating RTL as a flat token sequence obscures hardware dataflows, whereas exposing complete synthesized graphs overwhelms the model with fine-grained metadata. To resolve this representation dilemma, we propose GRODE-RTL, an anchor-centric representation that extracts the single-clock-stage fan-in cone of a suspect register from the netlist, maps netlist nodes back to editable source lines, and organizes them into a compact hop-partitioned dependency view. By prioritizing direct drivers and folding distant hops, GRODE-RTL guides standard tool-using agents without exposing ground-truth hints or oracle feedback. We evaluate GRODE-RTL across 30 formally validated non-local faults using three LLMs (DeepSeek-V4-Flash, GLM-5.2, and MiniMax-M3) alongside a controlled five-configuration ablation study. GRODE-RTL improves pass@1 over source-level exploration by 16.7\% on DeepSeek, by 26.7\% on GLM-5.2, and by 20.0\% on MiniMax-M3, with consistent gains on the hard-to-retrieve far-distance and indirect-anchor subsets of all three models. Our controls isolate structural relevance from generic context expansion, proving that selective dependency organization rather than full graph exposure is key to effective RTL repair.
\end{abstract}

\section{Introduction}

LLM-based agents debug software with human-level competence, yet this success only partially transfers to Register-Transfer Level (RTL) designs due to \emph{non-local faults}. In RTL, a divergent output and its offending statement often lie hundreds of lines apart, separated across module boundaries, deeply nested conditional chains, or long dependency paths. While local bugs confined to narrow code windows are easily resolved by LLMs, hardware concurrency and pipelining mean the vast majority of real-world RTL bugs are inherently non-local. Resolving this dominant fault class is thus essential for viable automated hardware repair.

Repairing a non-local RTL fault requires bridging a long causal gap from
an observable failure to the source statement that must be edited. This
process involves three distinct stages: localizing a suspicious signal
from the failing behavior~\citep{sepidband2026flcontext, sepidband2026rgfl}, reconstructing the static dependencies from
that signal to its potential source-level causes, and generating and
validating a patch. Existing repair methods primarily strengthen the two
ends of this process. Feedback-driven approaches use compiler, simulation,
or sequential-behavior feedback to generate and refine candidate
patches~\citep{tsai2022cirfix, du2026seqfeed,blocklove2023autochip}, while trace-based approaches
use execution behavior to narrow failures to suspicious signals or
registers~\citep{wang2026veripilot}.

\begin{figure}[t]
    \centering
    \includegraphics[width=0.95\linewidth]{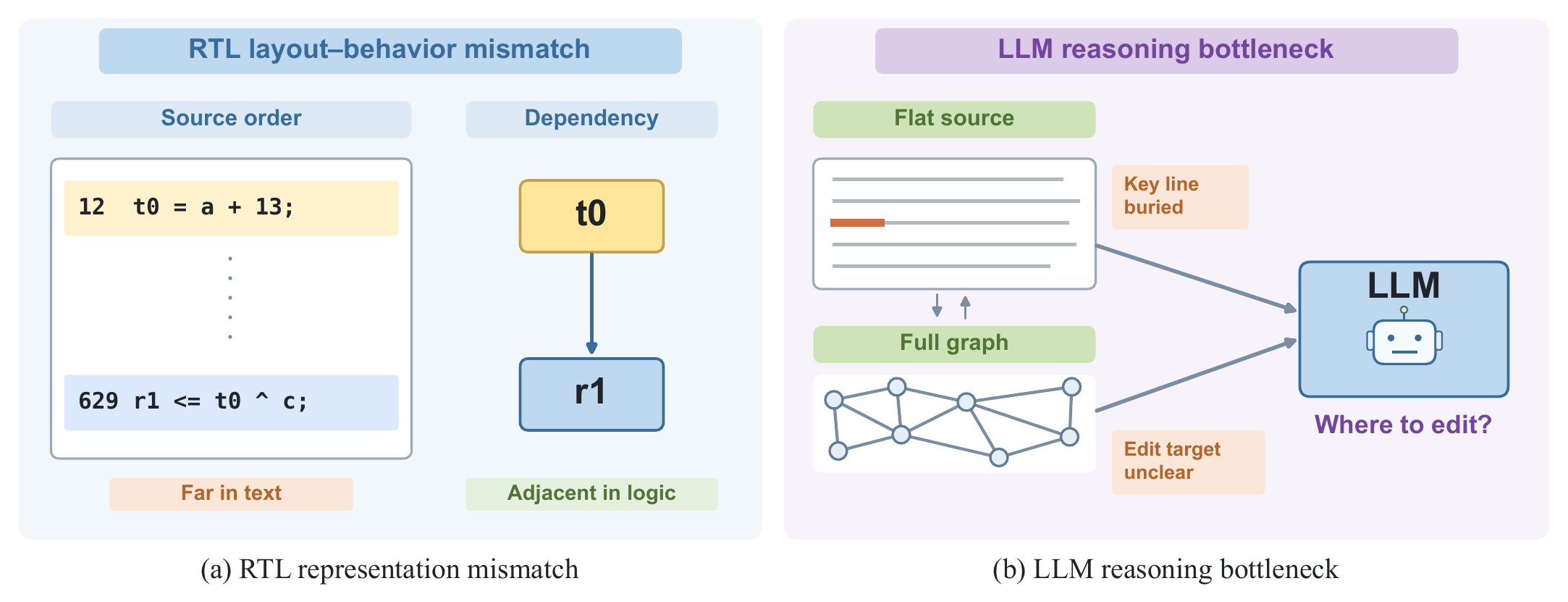}
    \caption{The representation dilemma for LLM-based non-local RTL repair. \textbf{Left:} Flat source obscures multi-hop dataflow dependencies across physical lines. \textbf{Right:} Passing raw netlists burdens LLMs with metadata, whereas GRODE-RTL extracts, organizes, and compresses fan-in logic into a source-mapped, compact view.}
    \label{fig:problem}
\end{figure}

However, neither line of work explicitly constructs the intermediate
repair context required by non-local faults. Sequential feedback reveals
whether and when candidate behavior diverges, but does not expose which
distant source statements statically influence the suspicious signal.
Signal-level localization identifies an anchor, but still leaves the
agent to search across modules, conditional chains, and dependency paths
for an editable root cause. The agent must therefore bridge behavioral
evidence and source-level repair locations using either flat RTL text or
a raw netlist, leading to a representation dilemma (Figure \ref{fig:problem}): (1)
 \textbf{Layout--Behavior Mismatch in Flat Source.}
RTL text is organized by layout rather than dataflow. Expanding contiguous source windows to capture distant drivers introduces irrelevant code, exacerbating needle-in-a-haystack localization problems.
(2) \textbf{The Graph Representation Dilemma.}
Exposing full netlists captures dependencies but inflates context with low-level metadata, obscuring editable source mappings and repair locations.

To bridge this gap, we propose GRODE-RTL, an anchor-centric, repair-oriented representation constructed via three key operations: (1)~\emph{Relevant code filtering} extracts the single-clock-stage fan-in cone of a suspicious anchor and maps netlist nodes back to editable RTL statements; (2)~\emph{Explicit dependency organization} orders statements by reverse-dependency depth and tags \texttt{DATA}, \texttt{CONTROL}, and \texttt{STATE} relations; and (3)~\emph{Selective graph compression} caps per-hop nodes and folds distant layers into unfoldable summaries.

Our main contributions are:
\begin{itemize}
    \item We expose the representation dilemma in non-local RTL repair, where flat source code lacks distant context while complete graphs overload LLM attention.
    \item We present GRODE-RTL, a representation pipeline that combines \textbf{source-mapped cone extraction} to eliminate design noise with \textbf{compact graph reorganization} to focus LLM reasoning.
    \item We demonstrate Pass@1 gains of up to \textbf{+26.7\,pp} across three LLMs (reaching a \textbf{63.3\%} success rate), with ablations confirming that cone extraction drives the primary performance leap while structural compression optimizes efficiency and edit precision.
\end{itemize}
\section{Related Work}
\label{sec:related-work}

\paragraph{Automated Repair for Hardware Designs.}
Automated hardware description language (HDL) repair originated with heuristic Search-Based Program Repair techniques such as CirFix~\citep{tsai2022cirfix}, which applies genetic programming over line-level coverage. With the emergence of Large Language Models, LLM-based RTL repair frameworks have advanced from single-prompt repair and conversation-driven refinement like AutoChip~\citep{blocklove2023autochip} to autonomous, tool-augmented debugging agents such as VeriPilot~\citep{wang2026veripilot} and SeqFeed~\citep{du2026seqfeed}. However, SeqFeed's sequential feedback (SeQuery and SeGraph) and VeriPilot's golden-guided control-data-flow tracing expose temporal simulation behavior rather than reorganizing the static dependencies of the current buggy design. GRODE-RTL instead introduces an anchor-centric structural representation between the SAT witness and the editing tools, transforming formal netlist dependencies into an interactive, source-mapped view for the agent.

\paragraph{Program and Graph Representations for LLMs.}
Graph, dataflow, and control-flow abstractions are widely used to enrich program context for machine learning models, often via learned GNN embeddings~\citep{kipf2017semi,velickovic2018gat,graphsage,grannite}. While effective for general code understanding, generic program graphs fail to capture hardware-specific state boundaries and clock-stage constraints. GRODE-RTL leverages these hardware properties by partitioning netlist fan-in cones into reverse-dependency hops, providing a human-readable and model-editable text view. Furthermore, our controlled ablations explicitly separate structural cone selection from generic context expansion.

\paragraph{Hardware Diagnosis and Netlist Representation Learning.}
Cone-of-influence (COI) reduction is a standard structural pruning technique in formal verification to bound target logic~\citep{handbookbmc,kuchlin2000proving,goldberg2001using,velev2001effective}. Concurrently, circuit representation learning (CRI) methods such as DeepGate variants~\citep{li2022deepgate,shi2023deepgate2,zheng2025deepgate4,wang2025moss} embed netlist structures to predict testability or equivalence. However, classical COI operates strictly within formal solver engines, while neural netlist embeddings lack direct source-level mapping for code modification. GRODE-RTL bridges this gap by mapping netlist-level COI logic back to Verilog source statements, transforming formal structural reductions into an interactive context for LLM repair agents.
\section{Method}
\label{sec:method}

\begin{figure}[t]
    \centering
    \includegraphics[width=\linewidth]{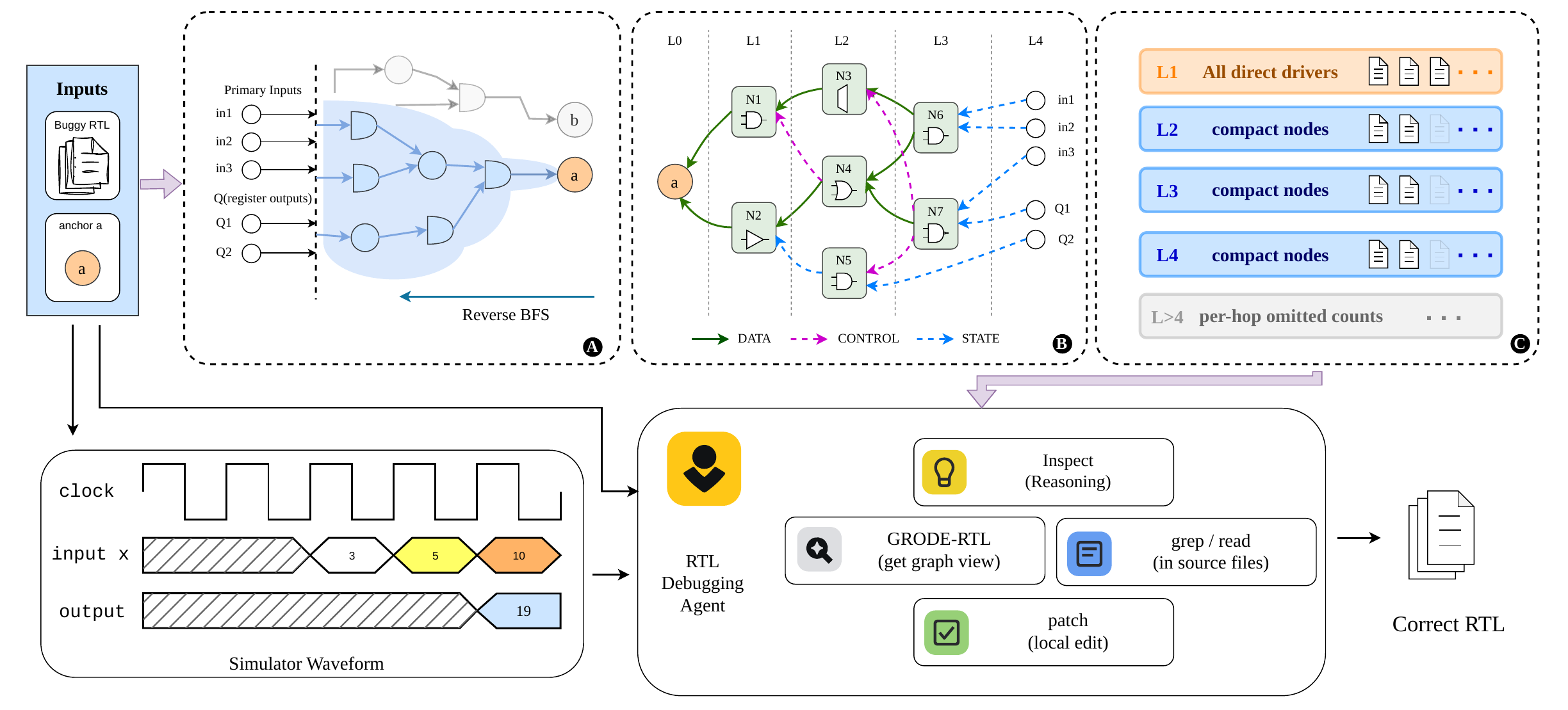}
    \caption{Overview of GRODE-RTL. \textbf{A) Subgraph Extraction} Synthesize buggy RTL and run reverse BFS from the anchor to recover the fan-in cone. \textbf{B) Dependency Organization} Partition the cone into per-hop levels $L_0, L_1, \dots$ by reverse-dependency depth. \textbf{C) Graph Compression} Select representative statements per hop and cap depth into a compact structural view.}
    \label{fig:overview}
\end{figure}

Figure~\ref{fig:overview} outlines the architecture of GRODE-RTL. 
To address unstructured source-level debugging, GRODE-RTL extracts the anchor's single-clock-stage fan-in cone from the synthesized netlist, organizes dependencies by reverse-depth, and compresses them into a compact, source-mapped structural view for the agent.

\subsection{Problem Formulation}
\label{sec:problem_formulation}
We define the problem of repairing a buggy RTL design. The agent receives the buggy Verilog source $P_b$, a counterexample waveform $\mathbf{x} \in \{0, 1\}^{L \times D}$ over $L$ clock cycles and $D$-bit bus width where $P_b$ diverges, and a suspicious anchor register $a$. The anchor $a = f_{\text{sim}}(P_b, \mathbf{x})$ is deterministically extracted as the first upstream divergent sequential register during counterexample simulation, providing the entry point for structural analysis. The counterexample and anchor serve as behavioral evidence, allowing the agent to observe where the divergence manifests without revealing the underlying root cause. Within a budget of $T$ turns, the agent modifies a range of at most 20 lines in $P_b$ to generate candidate $\hat{P}$ via \texttt{patch()}, submitting via \texttt{done()} or terminating via \texttt{abstain()}. The agent edits raw RTL source directly, never the netlist, and complete file rewrites are prohibited.

\subsection{Anchor-Centric Fan-In Extraction}

We synthesize $P_b$ via Yosys~\citep{wolf2013yosys} into a netlist $N = (\Gamma, W, \text{src})$ of cells $\Gamma$, wires $W$, and source locations $\text{src}: \Gamma \cup W \to \text{Locations}(P_b)$. From $N$, we extract the statement cone governing the anchor's next-state update.

\paragraph{State Boundary.}
Let $B \subseteq W$ be the set of Q-outputs of all sequential cells in $N$. To enforce a standard single-clock-stage abstraction, the fan-in cone is defined as the closure of $a$ under combinational reverse dependencies without crossing $B$. By bounding extraction at $B$, GRODE-RTL targets intra-stage combinational non-locality within a single clock cycle, deferring multi-cycle temporal unrolling to future work.

\paragraph{Cone Traversal.}
Defining one-step dependency $u \leadsto v \iff \exists c \in \Gamma, v \in \text{fanin}(c), u \in \text{fanout}(c)$, the anchor's fan-in cone $C(a)$ is:
\begin{equation}
    C(a) = \{w \in W \mid a \leadsto^{*} w, \text{ no element of } B \text{ on the path}\},
\end{equation}
where $\leadsto^{*}$ denotes the reflexive transitive closure. We compute $C(a)$ via a reverse BFS seeded at $a$, terminating at reachable Q-outputs and primary inputs.

\paragraph{Source Mapping.}
Using Yosys \texttt{src} annotations, we project the netlist cone $C(a)$ back to the unique RTL statements in $P_b$ that generated it:
\begin{equation}
    S(a) = \left\{ \text{src}(w) \mid w \in C(a) \right\} \subseteq \text{Lines}(P_b).
\end{equation}
Deduplication handles the many-to-one mapping from netlist elements to source lines (Appendix~\ref{app:src-recovery}).

\subsection{Cone Partition by Reverse Hop}

We partition $C(a)$ by reverse-dependency depth to construct a layered view of the computation cone, represented as a typed dependency graph $G = (V, E)$ with $V = V_N \cup V_S$: $V_N$ are the mapped source statements $S(a)$, and $V_S$ are the boundary signals where the reverse traversal terminates (sequential Q-outputs in $B$ and primary inputs).

\paragraph{Hop Partition.}
Let $h(v)$ denote the shortest directed path length from $a$ to $v$ in the reverse dependency graph. We partition the cone into per-hop layers $L_k = \{v \mid h(v) = k\}$ for $k = 0, 1, 2, \dots$, where $L_0 = \{a\}$, $L_1$ contains immediate drivers of $a$, $L_2$ drives $L_1$, and so forth. This unique partition is computed directly during the reverse BFS traversal.

\paragraph{Edge Typing.}
Edges in the dependency graph are assigned mutually exclusive labels $\tau \in \{\texttt{DATA}, \texttt{CONTROL}, \texttt{STATE}\}$ based on their operational roles: \texttt{DATA} for arithmetic and logic operands, \texttt{CONTROL} for multiplexer and conditional branching conditions, and \texttt{STATE} for dependencies anchored at sequential register boundaries ($B$). Appendix~\ref{app:dep-graph} details the cell-to-graph translation algorithm.

\subsection{Compact Graph View}
\begin{figure}[t]
    \centering
    \includegraphics[width=0.9\linewidth]{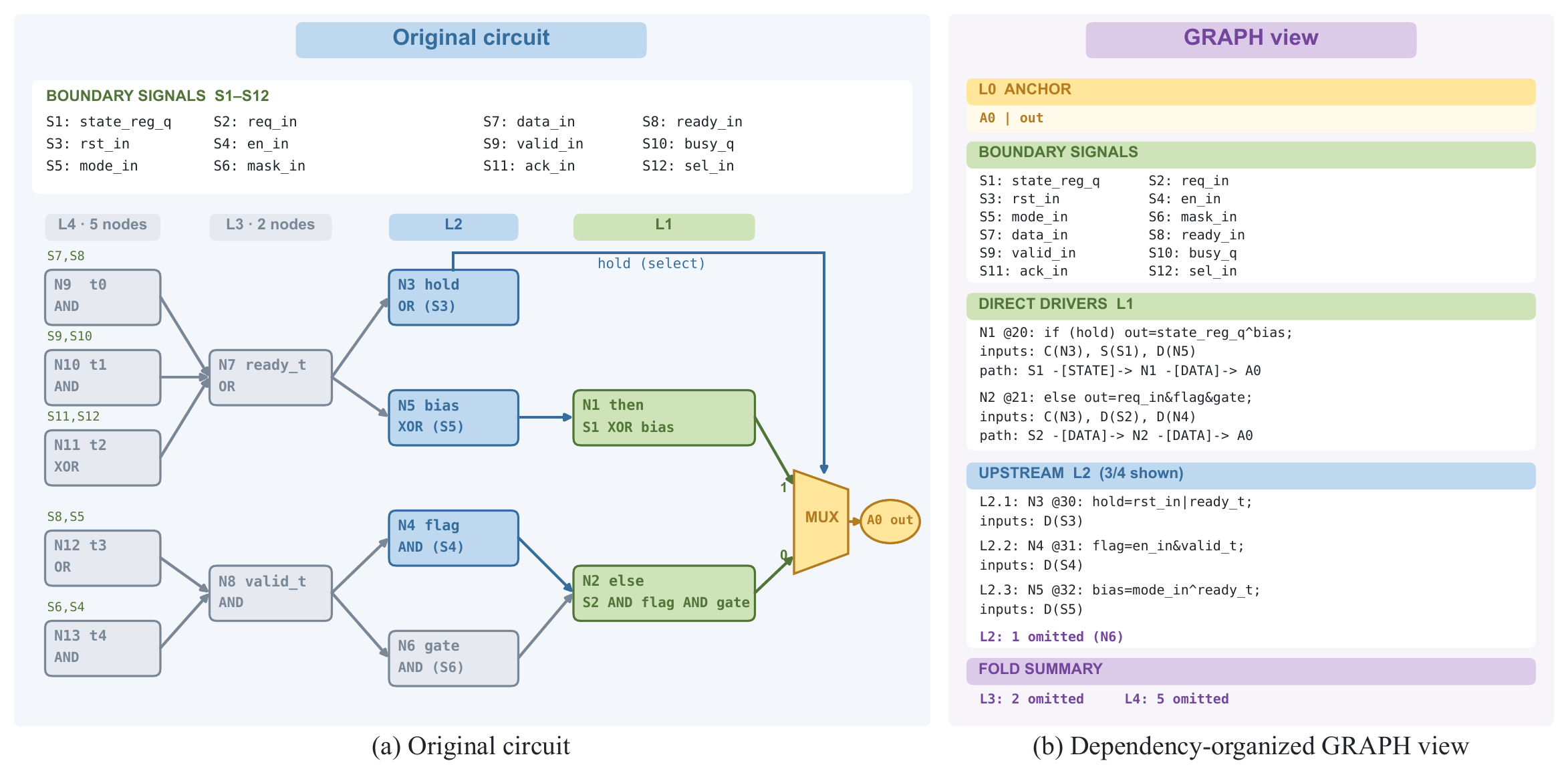}
    \caption{Anchor's single-clock-stage fan-in cone of a far-distance \texttt{const\_tweak} sample and the corresponding compact graph view. \textbf{Left} The cone recovered from Yosys. \textbf{Right} The compact graph view submitted to the agent.}
    \label{fig:graph-view}
\end{figure}
The compact graph view, serialized from $V$ and $E$, is submitted to the agent under four policies to bound token footprint while prioritizing structural attention (Figure~\ref{fig:graph-view}). By default, per-hop node cap $K = 4$ and visible depth cap $H = 4$.

\paragraph{Restructuring Policies.}
(1)~\textbf{Direct-driver priority} $L_1$ is emitted in full as the primary search space. 
(2)~\textbf{Per-hop representative cap} For $2 \le k \le H$, at most $K$ representative nodes are emitted from $L_k$ in reverse-BFS order. 
(3)~\textbf{Hop cap and summary folding} Layers beyond $L_H$ are folded into summaries such as \texttt{"L5: 8 omitted"}, acting as cues for dynamic unfolding via \texttt{read}. 
(4)~\textbf{Path retention} $L_1$ retains at least one shortest path to the state boundary $V_S$.
Formally, the view emits vertex set 
\begin{equation}
    V^{\text{compact}} = L_0 \cup L_1 \cup \bigcup_{k=2}^{H} \text{top}_{K}(L_k),
\end{equation}
fold summaries 
\begin{equation}
    F = \{(h, |L_h|) \mid h > H, L_h \neq \varnothing\},
\end{equation}
and induced typed edges $E^{\text{compact}}$.

\paragraph{Textual Serialization.}
$V^{\text{compact}}$, $E^{\text{compact}}$, and $F$ are serialized into an anchor-centric hierarchical text block. Each statement is annotated with line numbers, dependency types (\texttt{DATA}, \texttt{CONTROL}, \texttt{STATE}), and hop distance $h$, serving as a structured roadmap for precise edits.

\subsection{Agent Interaction and Information Boundary}

GRODE-RTL operates as an interactive representation layer enabling multi-turn, hop-aware localization and repair.

\paragraph{Tool-Augmented Repair Loop.}
The agent accesses standard tools (\texttt{read}, \texttt{grep}, \texttt{patch}, \texttt{done}, \texttt{abstain}) alongside GRODE-RTL's \texttt{get\_graph\_view()}, which returns $(V^{\text{compact}}, E^{\text{compact}}, F)$ on demand. Patch modifications are restricted to incremental, line-bounded edits to enforce surgical repairs.

\paragraph{Information Boundary.}
To prevent shortcut learning, the agent observes only the buggy source $P_b$, counterexample $\mathbf{x}$, anchor $a$, and structural graph views. Golden reference RTL, ground-truth fault locations, and internal oracle diagnostics are strictly concealed. Submissions are evaluated externally via sequential equivalence checking.
\section{Experiments}
\label{sec:experiments}
\definecolor{lightlightblue}{rgb}{0.91, 0.95, 0.98}
\subsection{Research Questions}

We organize the evaluation around three research questions.

\textbf{RQ1: Does GRODE-RTL improve LLM-based RTL repair?}
We evaluate whether the proposed structural augmentation improves repair
effectiveness across different LLMs and remains useful for bugs that are
difficult to retrieve through source-level exploration.

\textbf{RQ2: Why does the compact dependency representation help?}
We isolate structural relevance, dependency organization, and compactness
to determine whether the benefit comes from simply exposing more context
or from selectively presenting useful dependency information.

\textbf{RQ3: What are the behavioral, cost, and robustness trade-offs?}
We examine how structural guidance changes the agent's editing behavior,
what reasoning cost it introduces, and how timeout, statistical, and
fault-pattern effects bound the strength and scope of the conclusions.

\subsection{Experimental Setup}

\paragraph{Benchmark.}
Our evaluation builds on 85 single-site faults across 55 DeepCircuitX~\citep{li2025deepcircuitx} RTL designs, where each fault is verified non-equivalent under a 20-step zero-initialized SAT check~\citep{bmc2001} and lies within its anchor's fan-in cone. We evaluate 30 medium- and long-distance faults (23 designs) with source distance $d$ (anchor declaration to fault site) ranging from 50 to 782 lines (median: 104.5): 18 \emph{mid}-distance ($50 \le d \le 150$) and 12 \emph{far}-distance ($d > 150$). Designs range from 63 to 1,137 lines (median: 305). By anchor visibility, 16 are \emph{direct} (anchor name on faulty line) and 14 are \emph{indirect} (anchor name omitted). Fault types comprise 17 \texttt{signal\_swap}, 8 \texttt{const\_tweak}, and 5 \texttt{op\_swap}.

\paragraph{Common Protocol.}
All configurations share identical task prompts, buggy files, and SAT reproduction inputs, revealing neither golden RTL nor fault sites. Episodes run with temperature 1.0, maximum output $8{,}192$ tokens, master seed 0 ($k=1$), a 12-turn cap, 300~s episode timeout, and 600~s oracle timeout. Each episode permits up to 5 combined \texttt{grep}/\texttt{read} calls and 1 \texttt{patch} call ($\le 20$ lines; no full rewrites). An isolated budget mode ensures structural tools do not consume text-exploration quotas.

\paragraph{Evaluation Metrics \& Oracle.}
Candidates are evaluated via Yosys 20-step bounded SAT equivalence checking against the golden RTL oracle. We report Pass@1 (percentage of single-submission successful repairs) and absolute differences in percentage points (pp).

\paragraph{Evaluation Scope \& Controlled Anchor.}
To isolate repair-context quality from localization accuracy, we fix one
anchor register $a$ for each case and provide the same anchor to every
configuration. The anchor is defined offline as the first upstream
divergent sequential register identified through counterexample simulation
on the faulty design. Agents receive only the anchor identity; neither the
golden RTL nor the fault location is revealed. The evaluation is therefore
conditioned on a shared localization starting point and measures how
effectively each representation connects that anchor to its editable
source-level causes, rather than evaluating end-to-end fault localization.

\begin{figure}[h]
    \centering
    \includegraphics[width=0.8\linewidth]{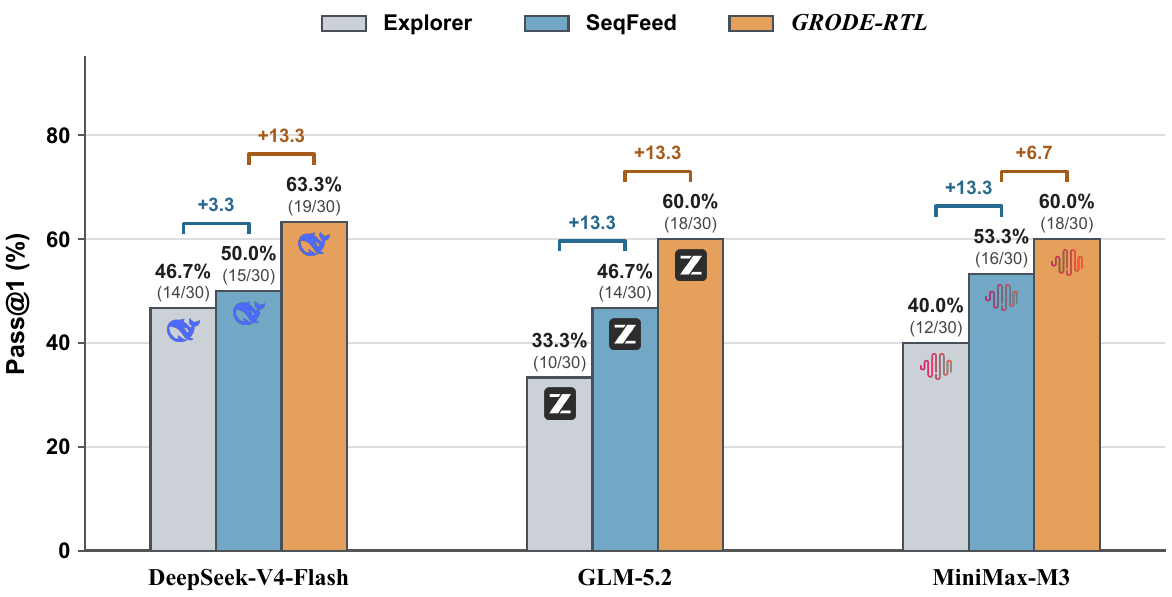}
    \caption{End-to-end pass@1 of Explorer, SeqFeed, and \emph{GRODE-RTL} on the
    same 30 mid- and far-distance bugs for each model.}
    \label{fig:effectiveness}
\end{figure}

\begin{table*}[h]
    \caption{Cross-model comparison of Explorer, SeqFeed, and GRODE-RTL on the 30 mid- and far-distance samples. }
    \label{tab:cross-model}
    \centering
    \small
    \renewcommand{\arraystretch}{1.05}
    \begin{tabular}{l l cccc}
        \toprule
        \multirow{2}{*}{\textbf{Model}} & \multirow{2}{*}{\textbf{Method}} & \multicolumn{4}{c}{\textbf{Pass@1 by subset}} \\
        \cmidrule(lr){3-6}
         &  & mid (18) & far (12) & direct (16) & indirect (14) \\
        \midrule
        DeepSeek-V4-Flash & Explorer & 9/18, 50.0\%  & 5/12, 41.7\%  & 10/16, 62.5\% & 4/14, 28.6\% \\
        DeepSeek-V4-Flash & SeqFeed           & 11/18, 61.1\% & 4/12, 33.3\%  & 11/16, 68.8\% & 4/14, 28.6\% \\
        \rowcolor{lightlightblue}
        DeepSeek-V4-Flash & \emph{GRODE-RTL}  & \textbf{11/18, 61.1\%} & \textbf{8/12, 66.7\%} & \textbf{11/16, 68.8\%} & \textbf{8/14, 57.1\%} \\
        GLM-5.2            & Explorer & 8/18, 44.4\%  & 2/12, 16.7\%  & 7/16, 43.8\%  & 3/14, 21.4\% \\
        GLM-5.2            & SeqFeed           & 10/18, 55.6\% & 4/12, 33.3\%  & 9/16, 56.2\%  & 5/14, 35.7\% \\
        \rowcolor{lightlightblue}
        GLM-5.2            & \emph{GRODE-RTL}  & \textbf{13/18, 72.2\%} & \textbf{5/12, 41.7\%} & \textbf{12/16, 75.0\%} & \textbf{6/14, 42.9\%} \\
        MiniMax-M3         & Explorer & 8/18, 44.4\%  & 4/12, 33.3\%  & 9/16, 56.2\%  & 3/14, 21.4\% \\
        MiniMax-M3         & SeqFeed           & 12/18, 66.7\% & 4/12, 33.3\%  & 11/16, 68.8\% & 5/14, 35.7\% \\
        \rowcolor{lightlightblue}
        MiniMax-M3         & \emph{GRODE-RTL}  & \textbf{12/18, 66.7\%} & \textbf{6/12, 50.0\%} & \textbf{12/16, 75.0\%} & \textbf{6/14, 42.9\%} \\

        \midrule
        \multirow{2}{*}{\textbf{Model}} & \multirow{2}{*}{\textbf{Method}} & \multicolumn{2}{c}{\textbf{Per episode (of 30)}} & \multicolumn{2}{c}{\textbf{Per-episode cost}} \\
        \cmidrule(lr){3-4}\cmidrule(lr){5-6}
         &  & Pass & No edit & \multicolumn{2}{c}{Tokens} \\

        \midrule
        DeepSeek-V4-Flash & Explorer & 14 & 12 & \multicolumn{2}{c}{55,982} \\
        DeepSeek-V4-Flash & SeqFeed           & 15 & 9  & \multicolumn{2}{c}{133,384}  \\
        \rowcolor{lightlightblue}
        DeepSeek-V4-Flash & \emph{GRODE-RTL}  & \textbf{19} & 11 & \multicolumn{2}{c}{72,317} \\
        GLM-5.2            & Explorer & 10 & 10 & \multicolumn{2}{c}{43{,}030}\\
        GLM-5.2            & SeqFeed           & 14 & 10 & \multicolumn{2}{c}{96,016} \\
        \rowcolor{lightlightblue}
        GLM-5.2            & \emph{GRODE-RTL}  & \textbf{18} & 5  & \multicolumn{2}{c}{57{,}774}\\
        MiniMax-M3         & Explorer & 12 & 10 & \multicolumn{2}{c}{63{,}054}  \\
        MiniMax-M3         & SeqFeed           & 16 & 7  & \multicolumn{2}{c}{147,426}  \\
        \rowcolor{lightlightblue}
        MiniMax-M3         & \emph{GRODE-RTL}  & \textbf{18} & 7  & \multicolumn{2}{c}{78,706}  \\
        \bottomrule
        
    \end{tabular}
\end{table*}

\subsection{RQ1: Does GRODE-RTL Improve LLM-based RTL Repair?}
\label{sec:effectiveness}

We evaluate three methods across 30 mid- and far-distance bugs ($3 \times 30$ evaluations) under the common protocol: source-only exploration (Explorer), the sequential-behavior-feedback baseline SeqFeed~\citep{du2026seqfeed}, and \emph{GRODE-RTL}.

\paragraph{Overall Effectiveness.}
\emph{GRODE-RTL} consistently boosts repair success across all evaluated LLMs (Figure~\ref{fig:effectiveness}, Table~\ref{tab:cross-model}). Across the three models, average pass@1 rises from 40.0\% (Explorer) to 50.0\% (SeqFeed), and reaches \textbf{61.1\%} with GRODE-RTL ($+21.1$~pp overall gain). The maximum improvement occurs on GLM-5.2 (33.3\% $\to$ 60.0\%, $+26.7$~pp), where GRODE-RTL nearly doubles the successfully repaired bugs (10 $\to$ 18). The consistent within-model gains validate that structural augmentation offers model-agnostic benefits to RTL repair workflows.

\paragraph{Dominance on Search-Hard Subsets.}
The advantage of \emph{GRODE-RTL} is especially pronounced on hard-to-retrieve bugs. On far-distance bugs ($d > 150$), average pass@1 jumps from 30.6\% to \textbf{52.8\%} (+22.2~pp); on indirect-anchor bugs (where fault lines omit the anchor name), success increases from 23.8\% to \textbf{47.6\%} (+23.8~pp). Notably, SeqFeed yields limited gains on these subsets (33.3\% for both): sequential feedback exposes temporal propagation, yet source-level search over long static dependency chains remains the agent's burden. In contrast, \emph{GRODE-RTL} successfully bridges distant computation paths even when source-level text search yields no immediate matches.

\paragraph{RQ1 Takeaway.}
GRODE-RTL provides substantial, model-agnostic repair gains over both raw exploration and execution-feedback baselines, demonstrating unique strength in resolving spatially obscured and distant RTL faults.


\subsection{RQ2: Why Does the Compact Dependency Representation Help?}
\label{sec:ablation}

\begin{figure}[h]
    \centering
    \includegraphics[width=0.75\linewidth]{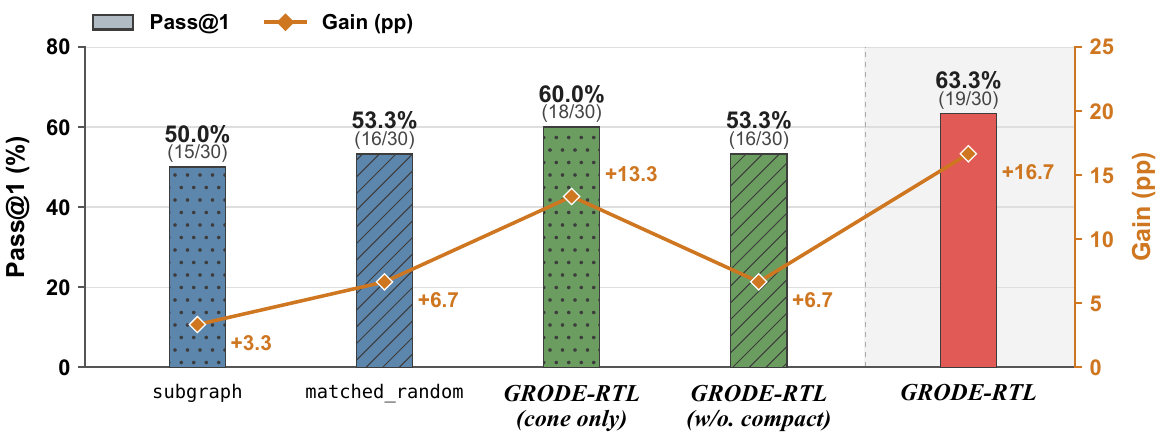}
    \caption{Pass@1 under different dependency representations on the same
    30 mid- and far-distance bugs.}
    \label{fig:ablation}
    \vspace{-10pt}
\end{figure}

To isolate the sources of improvement, we conduct a controlled representation ablation on DeepSeek-V4-Flash over the same 30 bugs (Table~\ref{tab:ablation-variants}). The control configurations incrementally incorporate structural features (Figure~\ref{fig:ablation}, Table~\ref{tab:configs}):
\begin{itemize}
    \item \texttt{subgraph}: Localized AST/dependency subgraph without source line context.
    \item \texttt{matched\_random}: Size-matched random subgraph to test functional relevance versus raw context volume.
    \item \emph{GRODE-RTL(cone only)}: Unorganized fan-in cone mapped to source statements without dependency hierarchy.
    \item \emph{GRODE-RTL(w/o. compact)}: Unpruned netlist dependency graph without selective cone partitioning.
\end{itemize}
Here, \textbf{view nodes} refers to the number of AST/dependency statement nodes serialized into the structural context, and \textbf{view chars} denotes its raw string length.
\definecolor{passgreen}{RGB}{0, 115, 38}    
\definecolor{failred}{RGB}{180, 40, 40}     
\newcommand{\greencheck}{\textcolor{passgreen}{\ding{51}}} 
\newcommand{\redx}{\textcolor{failred}{\ding{55}}}        

\begin{table}[h]
    \centering
    \caption{Structural features captured by representation variants in the ablation.}
    \label{tab:ablation-variants}
    \begingroup
    \footnotesize
    
    \setlength{\tabcolsep}{5pt}
    \renewcommand{\arraystretch}{1.15}

    \begin{tabular}{l c c c c c}
        \toprule
        \textbf{Configuration} &
        \begin{tabular}[c]{@{}c@{}}\textbf{Anchor}\\\textbf{Signal}\end{tabular} & 
        \begin{tabular}[c]{@{}c@{}}\textbf{Source}\\\textbf{Lines}\end{tabular} & 
        \begin{tabular}[c]{@{}c@{}}\textbf{Subgraph}\\\textbf{Extract}\end{tabular} & 
        \begin{tabular}[c]{@{}c@{}}\textbf{Dep.}\\\textbf{Org.}\end{tabular} & 
        \begin{tabular}[c]{@{}c@{}}\textbf{Graph}\\\textbf{Comp.}\end{tabular} \\
        \midrule
        \texttt{subgraph}       & \greencheck & \redx       & \redx       & \redx       & \redx       \\
        \texttt{matched\_random}& \greencheck & \greencheck & \redx       & \redx       & \redx       \\
        \emph{GRODE-RTL(cone only)}           & \greencheck & \greencheck & \greencheck & \redx       & \redx       \\
        \emph{GRODE-RTL(w/o. compact)}    & \greencheck & \greencheck & \greencheck & \greencheck & \redx       \\
        \rowcolor{lightlightblue}
        \emph{GRODE-RTL}        & \greencheck & \greencheck & \greencheck & \greencheck & \greencheck \\
        \bottomrule
    \end{tabular}
    \endgroup
\end{table}

\paragraph{Structural Relevance Over Raw Context.}
Exposing only the coarse SAT mismatch signal (tool-call \texttt{get\_subgraph()}) yields 50.0\% pass@1, just 3.3~pp above source-only exploration. Providing line-count-matched random lines (\texttt{matched\_random}) reaches 53.3\%, whereas supplying the actual fan-in cone (tool-call \texttt{get\_cone()}) raises pass@1 to 60.0\%. This confirms that structural relevance to the faulty computation---rather than coarse localization or generic context expansion---drives the performance gain.

\paragraph{Over-exposure Degrades Reasoning.}
Adding unpruned graph detail does not yield monotonic gains. While \texttt{full\_graph} introduces explicit hop partitions and typed edges, it drops overall Pass@1 to 53.3\% ($-6.7$\,pp), with severe drops on far-distance (58.3\% $\rightarrow$ 33.3\%) and indirect-anchor (50.0\% $\rightarrow$ 35.7\%) subsets. Excessive structural noise burdens LLM attention, counteracting the benefits of graph context.

\paragraph{Compactness and Typing Enable Targeted Exploration.}
Combinational fan-in slicing (\emph{GRODE-RTL(cone only)}) accounts for the primary recovery over baseline exploration (+13.3\,pp), matching \emph{GRODE-RTL} on simple direct-anchor (68.8\%) and mid-distance (61.1\%) subsets under a lower footprint. However, unorganized linear listings stall on complex far-distance and indirect-anchor faults. \emph{GRODE-RTL}'s typed hop-partitioning resolves this bottleneck, delivering vital incremental gains (+3.3\,pp overall) that reach peak performance on far-distance (66.7\%) and indirect (57.1\%) cases. Furthermore, compared to \emph{GRODE-RTL(w/o. compact)}, \emph{GRODE-RTL} prunes node count by 62.6\% (9.8 vs.\ 26.2) and view size by 47.0\% (2,398 vs.\ 4,524 chars), proving that selective, typed partitioning outperforms both unorganized linear slicing and unpruned context expansion.

\paragraph{RQ2 Takeaway.}
The ablation demonstrates a clear progression: localization offers minimal gain, raw dependency context helps, unpruned graph structures degrade reasoning, and GRODE-RTL optimizes the trade-off through compact dependency organization.

\begin{table*}[h]
    \caption{Five ablation configurations on DeepSeek-V4-Flash.}
    \label{tab:configs}
    \centering
    \scriptsize
    \setlength{\tabcolsep}{4pt}
    \renewcommand{\arraystretch}{1.05}
    \begin{tabular}{l cc cccc}
        \toprule
        & \multicolumn{2}{c}{\textbf{Overall}} & \multicolumn{4}{c}{\textbf{Pass@1 by subset}} \\
        \cmidrule(lr){2-3}\cmidrule(lr){4-7}
        Configuration & Pass & $\Delta$ & mid (18) & far (12) & direct (16) & indirect (14) \\
        \midrule
        
        \texttt{subgraph}       & 15/30, 50.0\%           & $-13.3\%$                 & 11/18, 61.1\% & 4/12, 33.3\%  & 11/16, 68.8\% & 4/14, 28.6\%  \\
        \texttt{matched\_random}& 16/30, 53.3\%           & $-10.0\%$                 & 11/18, 61.1\% & 5/12, 41.7\%  & 9/16, 56.2\%  & 7/14, 50.0\%  \\
        \emph{GRODE-RTL(cone only)}           & 18/30, 60.0\%           & ${-3.3\%}$         & 11/18, 61.1\% & 7/12, 58.3\%  & 11/16, 68.8\% & 7/14, 50.0\%  \\
        \emph{GRODE-RTL(w/o. compact)}    & 16/30, 53.3\%           & $-10.0\%$                 & \textbf{12/18, 66.7\%} & 4/12, 33.3\% & 11/16, 68.8\% & 5/14, 35.7\%  \\
        \rowcolor{lightlightblue}
        \emph{GRODE-RTL}        & \textbf{19/30, 63.3\%} & ---                       & 11/18, 61.1\% & \textbf{8/12, 66.7\%} & \textbf{11/16, 68.8\%} & \textbf{8/14, 57.1\%} \\
        \midrule
        & \multicolumn{2}{c}{\textbf{Edit (of 30)}} & \multicolumn{4}{c}{\textbf{Per-episode cost}} \\
        \cmidrule(lr){2-3}\cmidrule(lr){4-7}
        Configuration & Patches & No edit & Tool calls & Tokens & View nodes & View chars \\
        \midrule
        
        \texttt{subgraph}       & 20 & 9  & 5.73 & 63{,}235 & ---   & ---      \\
        \texttt{matched\_random}& 19 & 8  & 6.40 & 64{,}987 & 26.2 & 1{,}326 \\
        \emph{GRODE-RTL(cone only)}           & 21 & 9  & 6.57 & 65{,}987 & 24.7 & 1{,}761 \\
        \emph{GRODE-RTL(w/o. compact)}    & 17 & 12 & 6.03 & 70{,}112 & 26.2 & 4{,}524 \\
        \rowcolor{lightlightblue}
        \emph{GRODE-RTL}        & 19 & 11 & 6.43 & 72{,}317 & 9.8  & 2{,}398 \\
        \bottomrule
    \end{tabular}
    \vspace{2mm}
    \begin{minipage}{0.95\textwidth}
        {\tiny \textdagger~For configurations where Patches and No Edit sum to less than 30 (e.g., \texttt{subgraph}, \texttt{matched\_random}, and \texttt{full\_graph}), the remaining episodes terminated via timeout prior to issuing a patch call or exit decision.}
    \end{minipage}
\end{table*}

\subsection{RQ3: Behavioral, Cost, and Robustness Trade-offs}
\label{sec:tradeoffs}

RQ3 examines how structural guidance alters agent execution, its associated costs, and the bounds of our findings.

\paragraph{Selective Guidance Drives Precise Edits.}
As shown in Table~\ref{tab:cross-model}, GRODE-RTL consistently suppresses no-edit fallbacks across models (DeepSeek: 12$\to$11; GLM-5.2: 10$\to$5; MiniMax-M3: 10$\to$7) while increasing repair yield. Conditioned on edited episodes, conditional success rates surge: 77.8\%$\to$100.0\% (DeepSeek), 50.0\%$\to$72.0\% (GLM-5.2), and 60.0\%$\to$78.3\% (MiniMax-M3). This proves the representation enhances repair quality rather than mere edit propensity.

\paragraph{Reasoning Cost vs. Selective Exposure.}
Although GRODE-RTL increases per-episode token consumption compared to the baseline \textit{explorer} (e.g., 72,317 vs. 55,982 on DeepSeek-V4-Flash), it achieves substantial token savings compared to SeqFeed (72,317 vs. 133,384 tokens, a 45.8\% reduction; and a 39.8\%--46.6\% reduction across all backbones, e.g., 57,774 vs. 96,016 on GLM-5.2 and 78,705 vs. 147,426 on MiniMax-M3). This demonstrates that compact view synthesis successfully avoids context overload, enabling selective exposure to channel the token budget into targeted follow-up reads and deeper sequential reasoning.

\paragraph{Fault Observability and Evaluation Bounds.}
The structural gain varies by fault pattern: GRODE-RTL substantially lifts fully observable \texttt{const\_tweak} bugs (2/8$\to$6/8 on DeepSeek) but provides marginal gains on \texttt{signal\_swap} (12/17$\to$13/17) and fails on unobservable \texttt{op\_swap} faults (0/5), which demand semantic rather than structural reasoning. Additionally, while relay timeouts introduce operational noise, the qualitative case (Appendix~\ref{app:case}) confirms that compact views avoid the 27k-char view clutter that causes \texttt{full\_graph} to forfeit edits.

\paragraph{RQ3 Takeaway.}
GRODE-RTL converts idling into high-precision edits by channeling tokens into deep sequential reasoning rather than raw context display. Its strength centers on structural observability, establishing selective dependency organization as a robust principle within explicit operational boundaries.
\begin{revision}
\subsection{Repeated-run evaluation}
\label{sec:repeated}

We repeat the evaluation on DeepSeek-V4-Flash, GLM-5.2,
and MiniMax-M3 using seeds 101, 102, and 103.
Each run covers the same 30 faults from 23 source designs
and compares Explorer, \emph{GRODE-RTL(cone only)}, and \emph{GRODE-RTL},
yielding 810 episodes in total.
We use temperature 1.0 with a budget of 12 model calls
per episode and 8,192 output tokens per response.
Thinking is enabled for DeepSeek-V4-Flash.
Table~\ref{tab:repeated} summarizes the successful repairs
for each seed and reports mean pass@1 with the sample
standard deviation across the three runs.

\begin{table*}[t]
    \caption{Repeated-run repair results on the same 30 cases.
    Each seed column reports the number of successful repairs.
    Mean $\pm$ SD reports pass@1 in percent and its sample standard
    deviation across the three seeds. Best results within each model
    and column are bold.}
    \label{tab:repeated}
    \centering
    \color{black}
    \small
    \renewcommand{\arraystretch}{1.05}
    \begin{tabular}{l l cccc}
        \toprule
        \multirow{2}{*}{\textbf{Model}}
        & \multirow{2}{*}{\textbf{Method}}
        & \multicolumn{3}{c}{\textbf{Successful repairs (of 30)}}
        & \multirow{2}{*}{\textbf{Mean $\pm$ SD (\%)}} \\
        \cmidrule(lr){3-5}
        & & Seed 101 & Seed 102 & Seed 103 & \\
        \midrule
        DeepSeek-V4-Flash & Explorer
        & 19 & 14 & 20 & $58.9 \pm 10.7$ \\
        DeepSeek-V4-Flash & \emph{GRODE-RTL(cone only)}
        & \textbf{21} & \textbf{22} & 22
        & $\mathbf{72.2} \pm 1.9$ \\
        \rowcolor{lightlightblue}
        DeepSeek-V4-Flash & \emph{GRODE-RTL}
        & \textbf{21} & 21 & \textbf{23}
        & $\mathbf{72.2} \pm 3.8$ \\
        \midrule
        GLM-5.2 & Explorer
        & 10 & 11 & 15 & $40.0 \pm 8.8$ \\
        GLM-5.2 & \emph{GRODE-RTL(cone only)}
        & 17 & \textbf{19} & \textbf{19}
        & $\mathbf{61.1} \pm 3.8$ \\
        \rowcolor{lightlightblue}
        GLM-5.2 & \emph{GRODE-RTL}
        & \textbf{19} & 18 & 17 & $60.0 \pm 3.3$ \\
        \midrule
        MiniMax-M3 & Explorer
        & 1 & 0 & 2 & $3.3 \pm 3.3$ \\
        MiniMax-M3 & \emph{GRODE-RTL(cone only)}
        & 8 & 6 & \textbf{9} & $25.6 \pm 5.1$ \\
        \rowcolor{lightlightblue}
        MiniMax-M3 & \emph{GRODE-RTL}
        & \textbf{11} & \textbf{9} & \textbf{9}
        & $\mathbf{32.2} \pm 3.8$ \\
        \bottomrule
    \end{tabular}
\end{table*}

\paragraph{Repeated repair performance.}
Across all three evaluation backbones, \emph{GRODE-RTL} consistently outperforms Explorer across all three random seeds. On DeepSeek-V4-Flash, our method achieves a mean pass@1 of 72.2\% ± 3.8\%, compared to 58.9\% ± 10.7\% for Explorer. On GLM-5.2, the corresponding success rates are 60.0\% ± 3.3\% and 40.0\% ± 8.8\%, representing average improvements of 13.3 and 20.0 percentage points, respectively. The cone-only ablation performs exceptionally close to full \emph{GRODE-RTL} on both models, matching its mean on DeepSeek-V4-Flash and solving just one additional case across 90 total attempts on GLM-5.2.

On MiniMax-M3, \emph{GRODE-RTL} achieves a mean pass@1 of 32.2\% ± 3.8\%, outperforming \emph{GRODE-RTL(cone only)} at 25.6\% ± 5.1\% and Explorer at 3.3\% ± 3.3\%. Full \emph{GRODE-RTL} exceeds cone-only in two seeds and ties in the third, yielding six additional successful repairs across 90 attempts. This performance gain is especially pronounced on far-distance faults, where \emph{GRODE-RTL} succeeds in 15 out of 36 attempts versus 11 out of 36 for \emph{GRODE-RTL(cone only)} (an 11.1 percentage-point increase), compared to 14 out of 54 versus 12 out of 54 on mid-distance faults (a 3.7 percentage-point increase).

The lower absolute success rates on MiniMax-M3 stem primarily from frequent output truncations when long reasoning traces hit the 8,192-token limit. Whenever a truncated model response omitted a valid tool call, the evaluation harness triggered a finalization-only step, prematurely terminating the episode regardless of any unused interaction budget.

\paragraph{Core representation and compact organization.}
The repeated results highlight the contribution of the
structural context shared by \emph{GRODE-RTL(cone only)} and \emph{GRODE-RTL}.
\emph{GRODE-RTL(cone only)} retains the central cone-extraction and
source-mapping components, which connect circuit
dependencies to editable source statements.
\emph{GRODE-RTL} further organizes this context by prioritizing
direct drivers and grouping upstream dependencies.
The two configurations achieve similar repair rates on
DeepSeek and GLM, while the MiniMax results show an
additional gain from the complete configuration under
the evaluated settings, particularly on far-distance faults.

\end{revision}

\begingroup
\color{black}
\section{Conclusion}
\label{sec:conclusion}

We presented GRODE-RTL, an anchor-centric representation framework that bridges netlist dependencies with editable RTL source for non-local hardware repair. Across three LLMs, GRODE-RTL yields average Pass@1 gains of \textbf{+21.1\,pp} over exploration baselines and \textbf{+11.1\,pp} over strong sequential baselines (reaching \textbf{63.3\%} peak), demonstrating that selective cone partitioning outperforms exhaustive graph serialization. The compact representation enables deeper structural reasoning across extended debugging turns. Future work will extend GRODE-RTL to autonomous upstream localization and operator-level diagnosis.

\subsection*{Reproducibility statement}
The benchmark construction, structural extraction, agent evaluation, and formal-equivalence pipeline are described in Sections~\ref{sec:method} and~\ref{sec:experiments}. Further implementation details, full prompts, and per-case logs are provided in Appendix~\ref{sec:appendix}. All evaluation code, raw datasets, and reproduction scripts are included in the supplementary material.

\subsection*{AI use statement}
Generative AI tools were employed for manuscript drafting and language refinement. All experimental data, tables, and reported pass@1 metrics were cross-referenced with raw execution logs, which are preserved in the supplementary material for reproducibility. The authors take full responsibility for the final publication.

\bibliographystyle{iclr2027_conference}
\bibliography{iclr2027_conference}
\endgroup
\appendix
\section{Supplementary Material}
\label{sec:appendix}

\definecolor{passgreen}{RGB}{0, 115, 38}    
\definecolor{failred}{RGB}{180, 40, 40}     
\definecolor{neutralsub}{RGB}{110, 110, 110}

\newcommand{\pass}{\textbf{\textcolor{passgreen}{PASS}}}
\newcommand{\fail}[1]{\textcolor{failred}{\texttt{#1}}}
\newcommand{\neut}[1]{\textcolor{neutralsub}{\texttt{#1}}}

This appendix first reports the per-case outcomes across all evaluated
models and repair methods. It then \rev{illustrates the recorded behavior of compact and full views on a concrete sample}, makes the information boundary
between configurations explicit by listing the full prompt and tool
descriptions, and shows one representative episode under both the compact
and full-view configurations. We refer to the case-study sample as
\emph{Sample~S} throughout; only its filename is anonymized to
\texttt{<design>.v}.

\subsection{Per-case outcomes across models}
\label{app:per-case-results}

Table~\ref{tab:per-case-results} lists all 30 evaluation cases using the
original circuit name and injected fault type. For each model, it compares
\texttt{explorer}, SeqFeed, and GRODE-RTL
(\texttt{explorer\_graph\_view\_compact}). \textbf{PASS} denotes a correct
repair. For failures, \texttt{NE}, \texttt{TO}, \texttt{ECF}, and
\texttt{ENE} denote \texttt{no\_edit}, \texttt{timeout},
\texttt{equivalence\_check\_failed}, and
\texttt{edited\_not\_equivalent}.

\rev{These are mutually exclusive terminal outcome labels. In particular, \texttt{TO} may occur with or without a patch, and \texttt{NE} excludes a run assigned \texttt{TO}. Final-source change status is counted separately in Appendix~\ref{app:rebuttal-audit}. An \texttt{ECF} caused by verification timeout is unresolved rather than a proof of inequivalence.}

\begin{table*}[t]
\caption{Per-case outcomes across models and repair methods. All methods are evaluated on the same 30 cases for each model.}
\label{tab:per-case-results}
\centering
\tiny
\setlength{\tabcolsep}{2.4pt}
\renewcommand{\arraystretch}{0.95}
\resizebox{\textwidth}{!}{%
\begin{tabular}{@{}rll*{9}{c}@{}}
\toprule
\multirow{2}{*}{\#} & \multirow{2}{*}{Circuit} & \multirow{2}{*}{Fault type} 
& \multicolumn{3}{c}{\textbf{DeepSeek-V4-Flash}} 
& \multicolumn{3}{c}{\textbf{GLM-5.2}} 
& \multicolumn{3}{c}{\textbf{MiniMax-M3}} \\
\cmidrule(lr){4-6}\cmidrule(lr){7-9}\cmidrule(lr){10-12}
& & & Explorer & SeqFeed & GRODE & Explorer & SeqFeed & GRODE & Explorer & SeqFeed & GRODE \\
\midrule
01 & \texttt{14520\_2} & \texttt{signal\_swap} & \neut{NE} & \neut{NE} & \neut{NE} & \neut{NE} & \neut{NE} & \neut{NE} & \neut{TO} & \neut{NE} & \neut{NE} \\
02 & \texttt{14520\_2} & \texttt{op\_swap}     & \neut{NE} & \neut{NE} & \neut{NE} & \neut{NE} & \neut{TO} & \neut{TO} & \fail{ENE} & \fail{ENE} & \neut{NE} \\
03 & \texttt{1542}    & \texttt{const\_tweak} & \fail{ENE} & \fail{ENE} & \pass & \neut{NE} & \neut{NE} & \neut{NE} & \fail{ENE} & \neut{NE} & \pass \\
04 & \texttt{16017}   & \texttt{signal\_swap} & \pass & \pass & \pass & \pass & \pass & \pass & \pass & \pass & \pass \\
05 & \texttt{16331\_3} & \texttt{signal\_swap} & \neut{NE} & \pass & \pass & \neut{NE} & \neut{NE} & \pass & \neut{NE} & \pass & \pass \\
06 & \texttt{16331\_3} & \texttt{const\_tweak} & \fail{ENE} & \neut{NE} & \pass & \neut{NE} & \pass & \pass & \neut{NE} & \pass & \pass \\
07 & \texttt{1762\_14} & \texttt{const\_tweak} & \neut{NE} & \neut{NE} & \neut{NE} & \neut{TO} & \neut{TO} & \neut{TO} & \neut{NE} & \pass & \pass \\
08 & \texttt{21502}   & \texttt{const\_tweak} & \pass & \pass & \pass & \neut{TO} & \neut{NE} & \pass & \pass & \pass & \pass \\
09 & \texttt{24583\_12}& \texttt{const\_tweak} & \pass & \pass & \pass & \neut{TO} & \pass & \pass & \neut{NE} & \pass & \pass \\
10 & \texttt{24583\_12}& \texttt{signal\_swap} & \pass & \pass & \pass & \pass & \pass & \neut{NE} & \neut{NE} & \pass & \pass \\
11 & \texttt{24940\_10}& \texttt{signal\_swap} & \pass & \pass & \pass & \neut{NE} & \neut{NE} & \pass & \pass & \fail{ECF} & \pass \\
12 & \texttt{24940\_10}& \texttt{signal\_swap} & \pass & \fail{ENE} & \pass & \neut{TO} & \neut{NE} & \neut{NE} & \pass & \neut{NE} & \pass \\
13 & \texttt{25631}   & \texttt{signal\_swap} & \pass & \neut{TO} & \pass & \pass & \neut{NE} & \pass & \pass & \pass & \pass \\
14 & \texttt{25631}   & \texttt{signal\_swap} & \pass & \pass & \neut{NE} & \pass & \pass & \pass & \pass & \pass & \fail{ECF} \\
15 & \texttt{29019\_6} & \texttt{signal\_swap} & \fail{ENE} & \fail{ENE} & \neut{NE} & \neut{TO} & \neut{NE} & \neut{TO} & \fail{ENE} & \fail{ENE} & \fail{ECF} \\
16 & \texttt{300\_2}   & \texttt{signal\_swap} & \neut{NE} & \pass & \pass & \neut{NE} & \pass & \pass & \fail{ECF} & \pass & \neut{NE} \\
17 & \texttt{30095\_15}& \texttt{signal\_swap} & \pass & \pass & \pass & \pass & \pass & \pass & \pass & \pass & \pass \\
18 & \texttt{30895}   & \texttt{signal\_swap} & \pass & \neut{NE} & \pass & \neut{TO} & \pass & \pass & \fail{ECF} & \fail{ENE} & \neut{TO} \\
19 & \texttt{3125\_6}  & \texttt{signal\_swap} & \pass & \pass & \pass & \neut{TO} & \pass & \pass & \pass & \neut{NE} & \fail{ECF} \\
20 & \texttt{33437\_14}& \texttt{signal\_swap} & \fail{ENE} & \pass & \pass & \pass & \pass & \neut{TO} & \pass & \neut{NE} & \fail{ENE} \\
21 & \texttt{33437\_14}& \texttt{signal\_swap} & \pass & \pass & \pass & \pass & \pass & \pass & \neut{TO} & \fail{ECF} & \pass \\
22 & \texttt{33831\_8} & \texttt{signal\_swap} & \pass & \pass & \pass & \pass & \pass & \pass & \neut{NE} & \fail{ECF} & \pass \\
23 & \texttt{35009}   & \texttt{op\_swap}     & \neut{NE} & \neut{NE} & \neut{NE} & \neut{TO} & \neut{TO} & \neut{TO} & \pass & \pass & \pass \\
24 & \texttt{3629\_1}  & \texttt{const\_tweak} & \neut{NE} & \fail{ENE} & \neut{NE} & \pass & \neut{NE} & \pass & \pass & \pass & \pass \\
25 & \texttt{3629\_1}  & \texttt{op\_swap}     & \neut{NE} & \neut{NE} & \neut{NE} & \neut{NE} & \neut{TO} & \pass & \neut{NE} & \neut{NE} & \neut{TO} \\
26 & \texttt{6109\_13} & \texttt{const\_tweak} & \neut{NE} & \pass & \pass & \neut{TO} & \pass & \pass & \fail{ENE} & \pass & \pass \\
27 & \texttt{6716\_2}  & \texttt{op\_swap}     & \neut{NE} & \neut{NE} & \neut{NE} & \neut{NE} & \neut{TO} & \neut{TO} & \fail{ECF} & \neut{NE} & \neut{TO} \\
28 & \texttt{6726\_6}  & \texttt{const\_tweak} & \neut{NE} & \pass & \pass & \neut{TO} & \pass & \pass & \neut{NE} & \pass & \pass \\
29 & \texttt{6777\_6}  & \texttt{signal\_swap} & \pass & \neut{NE} & \neut{NE} & \pass & \neut{NE} & \neut{TO} & \pass & \pass & \neut{NE} \\
30 & \texttt{8610}    & \texttt{op\_swap}     & \neut{NE} & \fail{ENE} & \neut{NE} & \neut{NE} & \fail{ENE} & \neut{NE} & \fail{ENE} & \fail{ENE} & \neut{NE} \\
\midrule
\multicolumn{3}{r}{\textbf{PASS total}}
& \textbf{14/30} & \textbf{15/30} & \textbf{\textcolor{passgreen}{19/30}}
& \textbf{10/30} & \textbf{14/30} & \textbf{\textcolor{passgreen}{18/30}}
& \textbf{12/30} & \textbf{16/30} & \textbf{\textcolor{passgreen}{18/30}} \\
\bottomrule
\end{tabular}%
}
\end{table*}

\subsection{Case study: a far-distance \texttt{const\_tweak} under five configurations}
\label{app:case}

\paragraph{Design.}
Sample~S is a small synchronous Verilog module \texttt{glyph\_addr} that
maps a 6-bit character input \texttt{glyph} together with a 4-bit
vertical counter \texttt{vcnt} to a 10-bit output \texttt{address}. The
body of the module is a single synchronous \texttt{always} block
containing a 24-branch \texttt{case (glyph)} statement; each case branch
holds an \texttt{if}\,/\,\texttt{else if} chain over \texttt{vcnt} that
updates the internal \texttt{cur\_add} register by a constant base
expression indexed by \texttt{vcnt}. The \texttt{cur\_add} register is then assigned to the
\texttt{address} output combinationally on the next line. The full source
is 784 lines; much of the body consists of near-identical \texttt{else if}
branches differing only in their \texttt{vcnt} literal and, in the buggy
version, in one constant term.

\paragraph{Anchor.}
The anchor is the 10-bit output register \texttt{address}, declared in
the module port list near the top of the file. It is an
\emph{indirect} anchor in our taxonomy. The anchor name \texttt{address}
does not appear on any \texttt{cur\_add} update line, so a naive
\texttt{grep("address")} only returns the port declaration and the
combinational assignment, neither of which is the bug site. The
\texttt{get\_suspect\_subgraph} tool's coarse-localizer signal
(simulation mismatch at step 20 on the \texttt{address} register, identifying the first upstream divergent sequential register) is the only reason the agent has to focus on this register.

\paragraph{Fault.}
The fault on Sample~S is a single-literal \texttt{const\_tweak} on
the \texttt{else if (vcnt == 'd10)} branch within one
\texttt{case (glyph)} branch: the constant \texttt{10'd151} should
read \texttt{10'd150}, the same base used by every other
\texttt{else if} branch and by the \texttt{rst} initialization. The
fault line lies $d = 620$ source lines below the anchor declaration,
putting Sample~S in the \texttt{far} bucket of the benchmark (distance
measured from the anchor declaration to the fault site). The right
edit is
therefore a one-character change in a 24-branch switch where every other
branch is correct, and where the same constant appears many times across
the file.

\paragraph{Configuration outcomes on this sample.}
The five available configuration outcomes on this sample are listed in
Table~\ref{tab:case-outcomes}. Failure labels are \texttt{no\_edit}
(the agent exhausts its turn budget without calling \texttt{patch}) and
\texttt{edited\_not\_equivalent} (the agent emits a patch whose
post-edit design is not functionally equivalent to the golden RTL).

\begin{table}[t]
    \caption{Per-configuration outcome on Sample~S under each of the five structural configurations.}
    \label{tab:case-outcomes}
    \centering
    \begin{tabular}{lll}
        \toprule
        Configuration & Outcome & Failure type \\
        \midrule
        \texttt{matched\_random}       & \textbf{PASS} & None \\
        \texttt{subgraph}              & FAIL & \texttt{edited\_not\_equivalent} \\
        \texttt{cone}                  & \textbf{PASS} & None \\
        \texttt{full\_graph}           & FAIL & \texttt{no\_edit} \\
        \emph{GRODE-RTL}               & \textbf{PASS} & None \\
        \bottomrule
    \end{tabular}
\end{table}

The full-graph configuration fails by \emph{giving up}: it consumes its exploration
budget on reads and greps without ever calling \texttt{patch}. The compact
configuration patches line 629 with a one-character change and is graded PASS.
Matched random context also passes on this sample, but the contrast of
interest for the representational claim is full $\rightarrow$ compact:
the two configurations see the same cone source set serialized in two ways, and
only the compact view converts the structure into an edit. The full
graph view is 27{,}871 characters of node/edge metadata; the compact
view is 2{,}968 characters of which roughly 1{,}300 are the navigation
header and fold summary, leaving about 1{,}700 characters of structured
source.
On this sample, more graph is worse.

\paragraph{The compact view the agent saw.}
The following is the exact \texttt{get\_graph\_view("address")} response
delivered to the agent when it called the tool:

\begin{tcolorbox}[
    enhanced, breakable,
    width=\linewidth,
    colback=black!3, colframe=black!55,
    boxrule=0.4pt, arc=1pt, outer arc=1pt,
    left=4pt, right=4pt, top=4pt, bottom=4pt,
    before upper=\color{black!85}\scriptsize,
    fontupper=\ttfamily\scriptsize]
\begin{Verbatim}[fontsize=\scriptsize]
GRODE-RTL Compact Graph View for anchor [address]
(784 total source lines in P_b; showing 11 representative statements):
Primary structural search space: inspect L1 direct-driver cards first;
expand beyond the view only if they cannot explain the failure.
Compact graph guide: showing 11/161 mapped source statements.
This is navigation, not a complete cone (bounded by K=4, H=4).
Folded statements by hop (F):
  L5: 15 omitted; L6: 6 omitted; L7: 10 omitted; ... L18: 9 omitted.
Use read() around a listed source location when full branch syntax is needed.

GRODE-RTL dependency view (current RTL only; no fault-line scores)
Top module: glyph_addr
How to use: inspect [L1 Direct driver statements] first; then read
a local range around a listed line if more syntax is needed.
Do not grep the anchor unless no mapped source node is shown.

[L0 Anchor]
A0 | address

[L1 Direct-driver statements: inspect and compare first]
N1 | <design>.v:9  | assign address = cur_add;
N2 | <design>.v:13 | if (rst) cur_add <= 10'd150;
     inputs: CONTROL(S2); DATA(N3)
     retained path: S2 -[CONTROL]-> N2 -[DATA]-> A0

[L2 Upstream statements (capped at K=4)]
N3 | <design>.v:16 | inputs: CONTROL(S1); CONTROL(N5); CONTROL(N7); DATA(N4)
                     case (glyph)

[L3 Upstream statements (capped at K=4)]
N4 | <design>.v:19 | inputs: CONTROL(S3); DATA(S1); DATA(S3); DATA(N8)
                     if (vcnt == 'd0) cur_add <= (10'd150+{glyph,4'b0000})+...
N5 | <design>.v:36 | inputs: CONTROL(S1)          end
N6 | <design>.v:39 | inputs: CONTROL(S3); DATA(S1); DATA(S3); DATA(N9)
                     if (vcnt == 'd0) cur_add <= (10'd150+{glyph,4'b0000})+...
N7 | <design>.v:56 | inputs: CONTROL(S1)          end

[L4 Upstream statements (capped at K=4)]
N8 | <design>.v:20 | inputs: CONTROL(S3); DATA(S1); DATA(S3)
                     else if (vcnt == 'd1) cur_add <= (10'd150+{glyph,4'b0000})+...
N9  | <design>.v:40 | inputs: CONTROL(S3); DATA(S1)
                     else if (vcnt == 'd1) cur_add <= (10'd150+...)
N10 | <design>.v:60 | inputs: CONTROL(S3); DATA(S1)
                     else if (vcnt == 'd1) cur_add <= (10'd150+...)
N11 | <design>.v:80 | inputs: CONTROL(S3); DATA(S1)
                     else if (vcnt == 'd1) cur_add <= (10'd150+...)

[State/Input boundary V_S]
S1 | glyph   S2 | rst   S3 | vcnt
\end{Verbatim}
\end{tcolorbox}

The compact graph view makes three structural properties visible that flat source code buries: 
(1) the anchor $a$ has a clear direct driver in $L_1$ (\texttt{N2 = <design>.v:13}); 
(2) intermediate hops $L_2$--$L_4$ are strictly capped at $K=4$ representative nodes, while the remaining 150 upstream statements across hops $L_5$--$L_{18}$ are cleanly folded into $F = \{(5, 15), (6, 6), \dots\}$ summaries; 
(3) the sequential state boundary $V_S$ isolates \texttt{glyph, rst, vcnt} as the sole clock-stage inputs. 
\rev{After three reads and two grep calls exhausted the exploration budget, the agent obtained the compact view and subsequently patched line 629, changing 151 to 150. No source read followed graph retrieval. The final source matched the golden reference, but the trace records no \texttt{done()} call; candidate correctness and submission-protocol completion are distinct.}

\paragraph{The full view the agent saw, in contrast.}
The same anchor serialized in the \texttt{full\_graph}
configuration contains 161 mapped source nodes and 612 typed edges, organized by
reverse hop from L1 to L18. The first 2{,}000 characters and the last
2{,}000 characters of that view are:

\begin{tcolorbox}[
    enhanced, breakable,
    width=\linewidth,
    colback=black!3, colframe=black!55,
    boxrule=0.4pt, arc=1pt, outer arc=1pt,
    left=4pt, right=4pt, top=4pt, bottom=4pt,
    before upper=\color{black!85}\scriptsize,
    fontupper=\ttfamily\scriptsize]
\begin{Verbatim}[fontsize=\scriptsize]
Graph view for address [full graph view]
(784 source lines; showing 161):
... [legend and L0 anchor as in the compact view] ...

[Direct driver statements: inspect and compare first]
N1 | <design>.v:9      assign address = cur_add;
N2 | <design>.v:13     if (rst) cur_add <= 10'd150;
    inputs: CONTROL(S2); DATA(N3)
    path: S2 -[CONTROL]-> N2 ; N2 -[DATA]-> A0

[L2 Remaining upstream source statements]
N3 | <design>.v:16 | inputs: CONTROL(S1); CONTROL(N11);
                     CONTROL(N13); CONTROL(N15); +16 more
                case (glyph)

[L3 Remaining upstream source statements]
N4  | <design>.v:19 | ... if (vcnt == 'd0) cur_add <= (10'd150+...) ...
N5  | <design>.v:36 | end
N6  | <design>.v:39 | ... if (vcnt == 'd0) cur_add <= (10'd150+...) ...
...
[continues through L18 with 9-10 near-identical statements per hop]

[L18 Remaining upstream source statements]
N153 | <design>.v:34  | ... else if (vcnt == 'd15) cur_add <= (10'd150+...) ...
N154 | <design>.v:54  | ... else if (vcnt == 'd15) cur_add <= (10'd150+...) ...
...
N161 | <design>.v:194 | ... else if (vcnt == 'd15) cur_add <= (10'd150+...) ...

[State/input boundary signals]
S1 | glyph   S2 | rst   S3 | vcnt
\end{Verbatim}
\end{tcolorbox}

The full view faithfully lists every hop, every node, every edge label. On
this sample the result is a 27{,}871-character payload with 161 nodes,
612 typed edges, and 18 reverse-dependency hops whose per-hop bodies are
near-identical \texttt{else if} lines differentiated only by \texttt{vcnt}
literals. The agent that received it performed four more exploration
calls (three greps and a targeted read of lines 760--784) without ever
producing a patch, and exhausted its 5-call exploration budget.

\rev{\paragraph{Interpretation.}
The compact and full views are derived from the same cone, with different amounts of displayed detail. The compact-view run produces a candidate matching golden, while the full-view run ends without an edit. Section~\ref{sec:repeated} reports the additional repeated experiments.}

\subsection{Recovering src-less statements}
\label{app:src-recovery}

The src-mapping step in Section~\ref{sec:method} is many-to-one
after deduplication, but a small number of source statements are
not reachable through any Yosys cell annotation at all. We
recover them by walking the cone and the anchor's declaration
sites directly:

\begin{itemize}
    \item \emph{Anchor bare assignments.} An \texttt{assign $a$ = expr;}
    in the RTL is folded into the netlist as a direct connection
    and produces no per-cell \texttt{src}; we recover it by
    scanning $P_b$ for assignments whose LHS is the anchor and
    adding them as hop-$1$ nodes tagged \texttt{DIRECT}.
    \item \emph{Direct signal assignments.} Initial-value assignments
    of $a$ (synchronous reset defaults inside the
    \texttt{always} block, module-port defaults) carry the same
    \texttt{src}-less status; we recover them by collecting every
    statement whose target is the anchor.
    \item \emph{Boundary declarations.} The anchor's declaration
    site in the module port or the \texttt{always}-block local
    list is not part of any cell's \texttt{src} and is recovered
    by parsing $P_b$'s port list and \texttt{always} headers.
\end{itemize}
All three classes are added as hop-$1$ nodes tagged \texttt{DIRECT}
so that the compact view's direct-driver priority (§4.4) treats
them on equal footing with cells that do carry \texttt{src}. The
extractor operates only on $P_b$ and never admits a statement on
the basis of a difference against the golden code; this boundary
is what keeps the compact view's representational gain attributable
to organization rather than to oracle leakage.

\subsection{Recovering the dependency graph}
\label{app:dep-graph}

The typed dependency graph of Section~\ref{sec:method} is
recovered cell-by-cell from the Yosys netlist and then lifted to
source-statement and signal nodes.

\paragraph{Vertex construction.}
The vertex set is $V = V_N \cup V_S$ with
$V_N = S(a) \setminus \{\text{anchor declaration}\}$ representing the mapped
source statements and $V_S$ representing the upstream signal alphabet reached
by the reverse BFS that does not cross the sequential boundary
$B \subseteq W$. The anchor itself sits as a singleton in $L_0$. Every other
vertex carries the deduplicated \texttt{src} annotation of the
cells that produced it.

\paragraph{Edge construction.}
Edges point from a statement to the statements it depends on, matching
the $u \leadsto v$ convention of Section~\ref{sec:method} and the
reverse-BFS hop levels; for readability, the serialized view prints
retained paths in dataflow direction (e.g.\ \texttt{S2 -[CONTROL]->
N2 -[DATA]-> a}).
For each combinational cell $c$ that survives the cone closure,
we record an edge from the statement node of $v = \text{fanout}(c)$
to the statement node of each $u \in \text{fanin}(c)$, then label the edge
according to $c$'s cell type:
\begin{itemize}
    \item \texttt{DATA}: data operand of a logic or arithmetic cell
        (e.g.\ \texttt{\$and}, \texttt{\$add}).
    \item \texttt{CONTROL}: selector input of a multiplexor or
        branch condition of a \texttt{\$mux}/\texttt{\$bmux}; these
        drive which operand is observed rather than carrying data.
    \item \texttt{STATE}: an edge whose source is a Q-output wire
        in $B$, i.e.\ the only carrier of state across clock
        stages.
\end{itemize}
A single Yosys cell may therefore contribute one or more edges
to $E$, each with its own label. The labels are mutually
exclusive and exhaustive: every edge in $E$ receives exactly one
of the three types, and the union of the three covers $E$. This
is the implementation of the edge-type claim in
Section~\ref{sec:method}.

\subsection{System prompt and tool descriptions}
\label{app:prompt}

\rev{The following reproduces the historical base template and configuration-specific additions. The base-template statement about no preselected suspect signal differs from the controlled-anchor description in the experimental setup; the literal prompt is retained here for traceability.}

\begin{Verbatim}[fontsize=\footnotesize]
You are fixing a bug in an RTL (Verilog) design. The design currently
fails a functional-equivalence check against its correct reference.
There is no source location or preselected suspect signal. You must
locate the root cause yourself.

Tools
- grep(pattern) regex-search the CURRENT design source; returns
  "file:line: text".
- read(file, start, end) return a numbered range of the current source.
- patch(file, start, end, content) replace a small 1-based inclusive
  line range in the CURRENT file. Use this instead of rewriting the
  complete file.
- done() submit the current files for grading.
- abstain(reason) explicitly stop when you cannot produce a defensible
  fix.

Method
Use the available evidence to inspect suspicious local operators,
constants, and signal sources before proposing broader redesigns. Trace
back from a divergent signal through the logic that drives it, patch only
the smallest line range justified by the evidence, then submit it. A fix
that changes the suspected logic but breaks other outputs will fail. Never
resend or reconstruct a complete source file. You MUST finish by calling
either done() or abstain(reason); never stop with an ordinary text response.
\end{Verbatim}

When the configuration registers a structural tool, we appends a short tool-specific block \emph{only for the tools that configuration has}.
The conditional append for the compact-graph configuration is:

\begin{Verbatim}[fontsize=\footnotesize]
Additional tools available in this episode:
- get_suspect_subgraph(): coarsely locate a suspect dependency subgraph
  and return its root signal together with SAT reproduction evidence.
- get_graph_view(anchor): inspect an anchor-centric CURRENT-RTL
  dependency view grouped by reverse hop with DATA, CONTROL, and STATE
  edges.

Tool-specific method:
Call get_suspect_subgraph() early, then use its root signal with the
structural tool available in this episode.
Inspect get_graph_view() direct-driver cards first, then read a small
local range only when syntax context is needed; do not reflexively grep
the anchor.
\end{Verbatim}

The same template, with \texttt{get\_graph\_view} replaced by
\texttt{get\_cone(anchor): map the anchor's single-stage fan-in cone back
to CURRENT RTL source lines}, is used for the cone configuration. The
\texttt{subgraph} configuration receives a reduced append containing
only \texttt{get\_suspect\_subgraph} (no view tool is registered).
The \texttt{matched\_random} configuration receives the cone block
unchanged. The matched-random condition is realized inside the tool
implementation, which returns \rev{size-matched source lines sampled outside the cone (not an exact token-count-matched AST subgraph)}; the
agent's prompt and tool specification are byte-identical to the
source-order cone configuration. Two configurations with identical
tool sets therefore produce byte-identical system prompts; this is
verified by the test suite.

The per-tool description that the model sees as a JSON-Schema tool spec
is also fixed. For \texttt{get\_graph\_view}, the description string is:

\begin{Verbatim}[fontsize=\footnotesize]
Return an anchor-centric reverse-dependency view of the current RTL.
Call it early. It lists direct driver statements of the anchor first,
with their immediate DATA, CONTROL, and STATE inputs and one
deterministic shortest upstream path. It then lists the remaining
same source statements by reverse hop: L1 directly drives the anchor
and larger L values are farther upstream. A0 denotes the anchor, N#
a source-statement node, and S# an upstream state/input boundary.
Node numbers are stable identifiers, not suspiciousness rankings.
Use read() around a listed line for more local syntax; avoid
grep(anchor) unless this view has no mapped source nodes. Treat the
displayed direct drivers as the primary structural search space before
broad global search. The view uses only current RTL and never uses
golden RTL, the fault location, or a suspiciousness score.
\end{Verbatim}

This is the literal information boundary: no golden code, no fault
line, no fault pattern, no suspiciousness score, no patch hint appears
anywhere in the prompt, the tool spec, or the tool output.

\subsection{One sample, two configurations}
\label{app:trace}

We follow Sample~S under the compact-graph configuration (PASS) and the
full-graph configuration (FAIL with \texttt{no\_edit}). Both
configurations consume the same per-episode budget: at most 5
\texttt{grep}/\texttt{read} calls, at most 1 \texttt{patch}, and at most
1 call to each structural tool.

\paragraph{Compact-graph configuration (PASS).}
\begin{enumerate}
    \item \textbf{turn~0} \texttt{get\_suspect\_subgraph()} $\rightarrow$
        anchor = \texttt{address}, mismatch at step 20.
    \item \textbf{turn~0} \texttt{read(<design>.v, 1, 80)} (5{,}649 chars).
    \item \textbf{turn~1} \texttt{read(<design>.v, 81, 260)} (15{,}211 chars).
    \item \textbf{turn~2} \texttt{read(<design>.v, 261, 560)} (25{,}394 chars).
        Three reads cover lines 1--560 of the 784-line file.
    \item \textbf{turn~3} \texttt{grep("10'd[0-9]+\textbackslash+|glyph,4'b[01]+|4'b[01]+,vcnt")}
        $\rightarrow$ 8{,}530 chars of repeated \texttt{cur\_add} lines.
    \item \textbf{turn~3} \texttt{grep("d15|d16|...|d6[0-3]")} $\rightarrow$
        8{,}470 chars. The 5-call exploration budget is now full.
    \item \textbf{turn~4} \texttt{get\_graph\_view("address")} $\rightarrow$
        2{,}968 chars (the compact view in
        Section~\ref{app:case}, fold summary + 11 nodes).
    \item \textbf{turn~4} \texttt{grep("10'd1[0-9][0-9]|glyph,4'b[\^{}0]|4'b[\^{}0],vcnt")}
        $\rightarrow$ budget exhausted message.
    \item \rev{A model request/response is logged at turn~5 without a tool call. The next recorded request is labeled turn~12 after a terminal-tool nudge. The log does not contain separate requests for turns~6--11.}
    \item \textbf{turn~12} \texttt{patch(<design>.v, 629, 629,
        "else if (vcnt == 'd10) cur\_add <=
        (10'd150+\{glyph,4'b0000\})+\{4'b0000,vcnt\};")} $\rightarrow$
        patched line 629, replacing \texttt{10'd151} with \texttt{10'd150}.
    \item \rev{No turn~13 or \texttt{done()} is recorded. The episode ends with \texttt{termination=unresolved} and \texttt{done=false}. External grading returns PASS because the final candidate exactly matches the golden source.}
\end{enumerate}

\rev{This trace documents seven completed tool calls, including three reads, two greps, one graph retrieval, and a final patch step indexed as 12 due to a separate completion phase. This index does not indicate thirteen cumulative model API calls.}

\paragraph{Full-graph configuration (FAIL, \texttt{no\_edit}).}
\begin{enumerate}
    \item \textbf{turn~0} \texttt{get\_suspect\_subgraph()} $\rightarrow$
        anchor = \texttt{address}.
    \item \textbf{turn~0} \texttt{grep("module|always|assign|glyph|vcnt")}
        $\rightarrow$ 8{,}130 chars of mixed hits.
    \item \textbf{turn~1} \texttt{get\_graph\_view("address")} $\rightarrow$
        27{,}871 chars (the full view).
    \item \textbf{turn~1} \texttt{read(<design>.v, 1, 60)} (3{,}977 chars).
    \item \textbf{turn~2} \texttt{grep("10'd|'d6[0-9]|default|endcase|endmodule")}
        $\rightarrow$ 8{,}470 chars.
    \item \textbf{turn~2} \texttt{read(<design>.v, 760, 784)} (2{,}192 chars).
        The agent is now reading the tail of the file, far from the folded
        L3 block that contains the fault.
    \item \textbf{turn~3} \texttt{grep("\^{}[ ]+'d[0-9]+:")} $\rightarrow$
        1{,}732 chars listing all 24 \texttt{case} branch headers.
    \item \textbf{turn~3} \texttt{grep("vcnt")} $\rightarrow$
        budget exhausted message.
    \item The agent never calls \texttt{patch}; the episode terminates
        with no edit. Oracle returns FAIL with \texttt{no\_edit}.
\end{enumerate}

\rev{This paired example records a successful compact-view candidate and a full-view no-edit outcome under the same anchor and exploration quotas.}

\begin{revision}
\subsection{Edit-status and timeout accounting}
\label{app:rebuttal-audit}
\begin{table}[htbp]
\centering\small
\caption{Patch status at model-service timeout in the original
Explorer/GRODE comparison. Counts are episodes, not tool calls.}
\label{tab:timeout-audit}
\begin{tabular}{llccc}
\toprule
Model & Method & Before patch & After patch & Total \\
\midrule
GLM-5.2 & Explorer & 9 & 1 & 10 \\
GLM-5.2 & GRODE-RTL & 7 & 0 & 7 \\
MiniMax-M3 & Explorer & 2 & 0 & 2 \\
MiniMax-M3 & GRODE-RTL & 3 & 0 & 3 \\
\bottomrule
\end{tabular}
\end{table}
\paragraph{Candidate correctness and submission status.}
Repair success is determined by the final candidate source.
The agent's termination status, including \texttt{done()},
abstention, or an unresolved ending, is recorded separately.
Conditional success in the main text is calculated as
successful repairs divided by $30$ minus the
\textit{No edit} count in Table~\ref{tab:cross-model}.
Timeout episodes remain in this denominator unless they
are included in that table's \textit{No edit} count.

\paragraph{Final-source changes.}
We also inspect whether the final source differs from the
buggy input, regardless of how the episode ends.
The numbers of unchanged candidates for Explorer, SeqFeed,
and GRODE-RTL are 12/10/11 on DeepSeek, 19/15/12 on GLM,
and 10/7/8 on MiniMax, respectively.
This source-based count differs from the terminal
\texttt{NE} label because an episode can end in a timeout
without changing the source.

For example, MiniMax GRODE-RTL has five \texttt{NE}
episodes and three timeouts, all with unchanged final source.
Two timeouts involve no patch attempt.
The third follows a rejected patch that leaves the source
unchanged.
This explains the distinction between the five terminal
\texttt{NE} labels, the seven \textit{No edit} entries
in Table~\ref{tab:cross-model}, and the eight unchanged
final candidates.

\paragraph{Timeouts in the original GLM runs.}
The original GLM experiments record 10 API timeouts for
Explorer and seven for GRODE-RTL.
Of the 11 cases repaired by GRODE-RTL but not Explorer,
six involve an Explorer timeout.
Service failures therefore contribute to the observed gap.

Nine Explorer timeouts occur before a successful patch-tool
call and one occurs after a patch.
All seven GRODE-RTL timeouts occur before a successful
patch-tool call.
None of these 17 episodes reaches a successful
\texttt{done()} submission.

\end{revision}

\end{document}